\documentclass[aps,prl,showpacs,twocolumn,superscriptaddress]{revtex4}
\usepackage{bm,color}
\usepackage{here}
\usepackage{graphicx}
\usepackage{amsmath}
\usepackage{color}
\usepackage{comment}
\usepackage[version=3]{mhchem}
\begin{document}
\title {Proposed controlled creation and manipulation of skyrmions with spin-orbit torque}
\author{Yuto Uwabo}
\affiliation{Department of Applied Physics, Waseda University, Okubo, Shinjuku-ku, Tokyo 169-8555, Japan}
\author{Ryo Fujimoto}
\affiliation{Department of Physics and Mathematics, Aoyama Gakuin University, Sagamihara, Kanagawa 229-8558, Japan}
\author{Kimimaro Yanai}
\affiliation{Department of Physics and Mathematics, Aoyama Gakuin University, Sagamihara, Kanagawa 229-8558, Japan}
\author{Masahito Mochizuki}
\affiliation{Department of Applied Physics, Waseda University, Okubo, Shinjuku-ku, Tokyo 169-8555, Japan}
\begin{abstract}
The physical mechanisms underlying current-driven skyrmion motion include the spin-transfer torque exerted by a spin-polarized horizontal electric current and the spin-orbit torque exerted by a perpendicular spin current. Each mechanism requires a specific sample geometry and structural configuration. Regarding current-induced skyrmion creation, skyrmions can be efficiently created at low current densities via spin-transfer torque when an electric current is applied to a nanotrack structure with a small notch. However, an effective and controlled method for skyrmion creation via spin-orbit torque in notched nanotracks has yet to be established. Here we theoretically propose a method for the creation, driving, and deletion of skyrmions in a three-terminal magnetic heterojunction with a notch. Our proposal offers valuable insights into the design of techniques for skyrmion creation and manipulation using spin-orbit torque, which is essential for technical applications of magnetic skyrmions as information carriers in next-generation spintronic memory devices.
\end{abstract}
\maketitle

\section{Introduction}
\begin{figure}[t]
\includegraphics[scale=1.0]{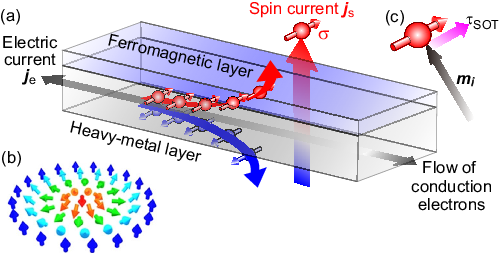}
\caption{(a) Ferromagnet/heavy-metal heterojunction. An electric current $\bm j_{\rm e}$ injected into the heavy-metal layer is converted into a vertical spin current $\bm j_{\rm s}$ with spin polarization $\bm \sigma$ via the spin Hall effect originating from interfacial spin-orbit interaction. This spin current exerts spin-orbit torque on the magnetization in the ferromagnetic layer. (b) Magnetization configuration of a N\'{e}el-type skyrmion. (c) Schematic illustration of the spin-orbit torque $\bm \tau_{\rm SOT}$. A vertical spin current with spin polarization $\bm \sigma_i$ exerts a torque on the local magnetization $\bm m_i$.}
\label{Fig01}
\end{figure}
Magnetic skyrmions are particle-like topological spin textures~\cite{Bogdanov1989,Bogdanov1994,Nagaosa2013} that emerge in various magnetic systems~\cite{SekiBook2016,Everschor2018,Tokura2020}, such as chiral magnets~\cite{Muhlbauer2009,YuXZ2010,Munzer2010,YuXZ2011,Seki2012a,Seki2012b,Tokunaga2015,Karube2017}, polar magnets~\cite{Kezsmarki2015,Fujima2017,Kurumaji2017,Kurumaji2021}, atomic layers~\cite{Romming2013,Heinze2011,Wiesendanger2016}, and magnetic bilayer heterojunctions~\cite{ChenG2015,Boulle2016,Moreau-Luchaire2016,Soumyanarayanan2017}. In these systems, the Dzyaloshinskii-Moriya (DM) interaction becomes active due to the broken spatial inversion symmetry~\cite{Dzyaloshinskii1957,Moriya1960a,Moriya1960b,Fert1980}. This interaction favors a rotating alignment of magnetizations with a pitch angle of 90$^\circ$ and competes with the ferromagnetic exchange interaction, which favors a parallel alignment of magnetizations. The strong competition between these interactions stabilizes a helimagnetic order with a moderate pitch angle. Upon the application of a magnetic field, this helimagnetic state transforms into a skyrmion crystal, in which skyrmions are periodically arranged to form a hexagonal lattice~\cite{Muhlbauer2009,YuXZ2010}.

In addition to the crystallized form, magnetic skyrmions can also appear as isolated topological defects in ferromagnets~\cite{YuXZ2010}. These individual skyrmions have been found to exhibit several advantageous properties for spintronic applications~\cite{Fert2013,Koshibae2015,Finocchio2016,KangW2016,Fert2017,XichaoZ2020}. Specifically, their potential use as information carriers in high-performance data storage devices has attracted significant attention, owing to their nanometric size, robustness due to topological protection, and extremely low energy cost for manipulation. These features enable the development of spintronic devices with high-density integration, resistance to external disturbances, and reduced power consumption. It is worth noting that skyrmion crystals can also emerge in centrosymmetric spin-charge coupled systems without DM interactions. In such systems, long-range skyrmion crystals arise from the frustration of long-range exchange interactions among localized spins, mediated by conduction electrons~\cite{Hayami2021R,Kawamura2025R}. The characteristic wavenumbers of these skyrmion crystals are determined by Fermi surface nesting. As a result, skyrmions in these systems always appear in a crystallized form rather than as individual defects.

One promising implementation of skyrmion-based magnetic memory devices is the so-called skyrmion racetrack memory~\cite{Tomasello2014,ZhangX2015,YuG2017,LaiP2017,Maccariello2018,ZhuD2018,HeB2023}, in which magnetic skyrmions replace the ferromagnetic domains used in conventional racetrack memory~\cite{Blasing2020,Parkin2015,Parkin2008}, which consists of ferromagnetic nanowires. A fundamental technology enabling skyrmion racetrack memory is the manipulation of skyrmions using electric currents~\cite{Fert2013,SeidelBook2016,Ohki2024,Everschor2012,Schulz2012}. This includes techniques for driving~\cite{Ohki2024,Jonietz2010,YuXZ2012,Iwasaki2013a,Iwasaki2013b,Sampaio2013,Iwasaki2014a,WooS2016,ZhangX2017,ZhangX2017b,Litzius2020,Reichhardt2022,ZhangX2023a,LuoJ2023,Pham2024}, creating~\cite{Iwasaki2013b,Sampaio2013,TchoeY2012,Yuan2016,YinG2016,Everschor-Sitte2017,Hrabec2017,YuXZ2017,YuXZ2020,Fujimoto2021,Fujimoto2022,WangW2022}, and deleting~\cite{Iwasaki2013b} skyrmions through current-induced magnetic torques acting on the noncollinear spin textures of skyrmions.

One key physical mechanism for this technique is the spin-transfer torque, which arises from the transfer of angular momentum from the spins of conduction electrons in the electric current to the noncollinear magnetization~\cite{Slonczewski1996,ZhangLi2004,Tatara2004}. The translational motion of skyrmions accompanied by a transverse deflection known as the skyrmion Hall effect~\cite{Iwasaki2013a,Iwasaki2013b,YangS2024,ZangJ2011,Everschor-Sitte2014,JiangW2017,Litzius2017} induced by spin-transfer torque has been extensively studied both theoretically and experimentally. Moreover, it has been theoretically predicted that individual skyrmions can be created by injecting an electric current into a magnetic nanostrip with a small notch via the spin-transfer torque mechanism. The theory has further suggested that skyrmions are successively created at the corner of the notch under specific conditions, which include the current density and direction, the sign of the external magnetic field, and the size of the notch~\cite{Iwasaki2013b}.

Skyrmion creation can be achieved through the local reversal of magnetization induced by an external stimulus, which usually requires a significant energy cost. The notch structure incorporated on a magnetic nanostrip helps to relax this topological constraint by introducing a discontinuity in the spatial distribution of magnetization. This discontinuity allows for a gradual rotation of magnetization vectors, rather than an abrupt flip, enabling the formation of a skyrmion core with magnetization oriented opposite to that of the surrounding area~\cite{Iwasaki2013b,YuXZ2020}. Furthermore, the notch facilitates the deterministic creation of skyrmions at a predetermined position, i.e., at a corner of the notch~\cite{Iwasaki2013b}. This clever method has been theoretically investigated not only for skyrmion creation with electric currents but also for creation using magnetic fields~\cite{Mochizuki2017}, electric fields~\cite{Mochizuki2015a,Mochizuki2015b}, and microwave fields~\cite{Miyake2020}.

In addition to spin-transfer torque, spin-orbit torque is another important physical mechanism~\cite{Manchon2019,ShaoQ2021,KimKW2024}. In a magnetic heterostructure where a ferromagnetic layer serves as the skyrmion host on top of a heavy-metal layer [Fig.\ref{Fig01}(a)], the broken spatial inversion symmetry at the interface gives rise to a strong spin-orbit interaction. This interfacial spin-orbit interaction also induces the DM interactions, which act on the magnetizations in the ferromagnetic layer and can stabilize magnetic skyrmions~\cite{ChenG2015,Moreau-Luchaire2016,Soumyanarayanan2017}. The skyrmions stabilized by this interfacial DM interaction are referred to as N'{e}el-type skyrmions and exhibit a fountain-like magnetization configuration, where the magnetizations rotate within the plane parallel to the radial direction [Fig.~\ref{Fig01}(b)]. When an electric current is injected into the heavy-metal layer, the interfacial spin-orbit interaction converts the horizontal charge current into a perpendicular pure spin current via the spin Hall effect, as shown in Fig.\ref{Fig01}(a). This spin current exerts a magnetic torque, known as spin-orbit torque, on the noncollinear magnetizations [Fig.\ref{Fig01}(c)]. It is well known that spin-orbit torque can drive and switch the magnetization more quickly and efficiently than spin-transfer torque~\cite{Miron2011,RyuKS2013,Emori2013}.

Current-driven motion of magnetic skyrmions via spin-orbit torque has been extensively studied and successfully demonstrated both theoretically and experimentally. However, skyrmion creation using spin-orbit torque has not yet been sufficiently understood~\cite{Buttner2017,LiuJ2022}. For practical applications, it would be highly beneficial if individual skyrmions could be created using spin-orbit torque through the simple injection of electric current into a magnetic bilayer, as has been demonstrated for spin-transfer torque. Recently, a simple current injection into a magnetic bilayer heterostructure with a notch was experimentally investigated~\cite{Buttner2017}. Although this study observed skyrmion creation, it failed to realize controlled or deterministic creation in terms of the dependence on current parameters and the precise location of skyrmion creation. This experiment underscores the necessity for further theoretical investigation into skyrmion creation via spin-orbit torque, both to refine the technique and to elucidate the underlying physical mechanisms.

In this paper, we theoretically investigate the controlled creation of individual skyrmions in a magnetic bilayer heterojunction with a small notch via the spin-orbit torque mechanism. We assume that a horizontal electric current injected to the heavy-metal layer is converted to a vertial spin current by the spin Hall effect due to the interfacial spin-orbit interaction and acts on the magnetization in the ferromagnetic layer via the spin-orbit torques. We first calculate the spatial distribution of current-density vectors of the injected electric current by numerically solving the Poisson equation using the finite-element method. Then we calculate the spatial distribution of the vertial spin current in the ferromagnetic layer on the basis of a natural assumption that the spin-current density is proportional to the electric current density, and the local spin polarization of the spin current is perpendicular to the local electric current. We perform micromagnetic simulations to investigate the spatiotemporal dynamics of magnetizations associated with the skyrmion creation via the spin-orbit torque mechanism. We find that individual skyrmions are indeed created in a controlled manner at a corner of the notch. Our proposal and finding will lead to an energy-saving and controlled method to create individual skyrmions in the magnetic bilayer system, which is anticipated to be a key technology for future spintronic applications of magnetic skyrmions.

In this paper, we theoretically investigate the controlled creation of individual skyrmions in a magnetic bilayer heterojunction with a small notch via the spin-orbit torque mechanism. We assume that a horizontal electric current injected into the heavy-metal layer is converted into a vertical spin current by the spin Hall effect caused by the interfacial spin-orbit interaction, and that this spin current exerts spin-orbit torque on the magnetization in the ferromagnetic layer. We first compute the spatial distribution of current density vectors by numerically solving the Poisson equation using the finite-element method. Based on the natural assumption that the spin-current density is proportional to the electric current density and that the local spin polarization is perpendicular to the local electric current, we then calculate the spatial distribution of the vertical spin current in the ferromagnetic layer. To investigate the spatiotemporal dynamics of magnetization associated with skyrmion creation via spin-orbit torque, we perform micromagnetic simulations, which reveal that individual skyrmions are indeed created in a controlled manner at the corner of the notch. Building on this, we theoretically propose a method for controlled creation, driving, and deletion of skyrmions in a three-terminal magnetic heterojunction with a notch. This proposal provides a promising route toward an energy-efficient and useful method for skyrmion operation in magnetic bilayer systems, which is expected to become a key technology for future spintronic applications of magnetic skyrmions.

\section{Model and Simulation}
\begin{figure}
\centering
\includegraphics[scale=1.0]{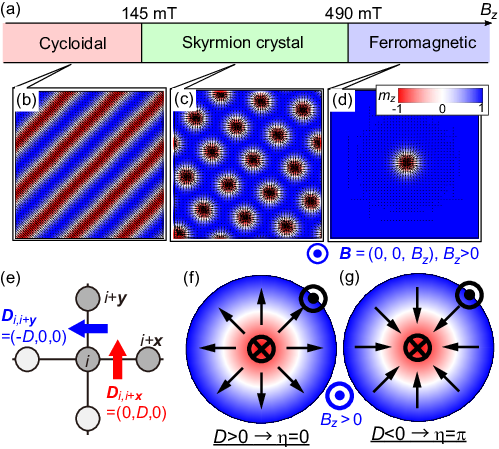}
\caption{(a) Ground-state phase diagram of the spin model in Eq.~(\ref{eq:model}) as a function of the perpendicular magnetic field $B_z$. (b)-(d) Magnetic structures of (b) cycloidal phase, (c) skyrmion crystal phase and (d) ferromagnetic phase including an isolated skyrmion as a topological defect. (e) Structure of the DM vectors which stabilizes the N\'{e}el-type skyrmion with the helicity $\eta=0$. (f),~(g) Schematics of the spatial magnetization configurations of the N\'{e}el-type skyrmions with different helicities $\eta$, i.e., (f) $\eta=0$ and (g) $\eta=\pi$, which depends on the sign of the DM parameter $D$.}
\label{Fig02}
\end{figure}
To describe the ferromagnetic layer of the magnetic bilayer heterojunction hosting N\'{e}el-type skyrmions, we employ a two-dimensional continuum spin model~\cite{Bak1980}, whose energy density is given by,
\begin{align}
\mathcal{E}
&=\mathcal{A}(\nabla \bm m)^2 - M_{\rm s}(\bm m \cdot \bm B_{\rm ext})
\nonumber \\
&-\mathcal{D} \left[
  \left(m_z \frac{\partial m_x}{\partial x} - m_x \frac{\partial m_z}{\partial x}\right)
+\left(m_z \frac{\partial m_y}{\partial y} - m_y \frac{\partial m_z}{\partial y}\right)
\right],
\label{eq:model}
\end{align}
This functional is composed of three terms which represent contributions from the ferromagnetic exchange interactions, the Zeeman interactions, and the DM interactions, respectively. Here $\bm m(\bm r)$ is the normalized magnetization vector at the spatial position $\bm r$. In the Zeeman-interaction term, we consider a magnetic field $\bm B_{\rm ext}=(0,0,B)$ applied perpendicular to the layer. For parameter values, we adopt $\mathcal{A}$=15 pJ\;m$^{-1}$ for the exchange stiffness, $\mathcal{D}$=3 mJ\;m$^{-2}$ for the DM parameter, and $M_{\rm s}$=480 kA\;m$^{-1}$ for the saturation magnetization.

Starting from this continuum model, we derive a classical Heisenberg model on a square lattice through dividing the continuum space into cubic cells with a volume of $a^3$. The Hamiltonian of this lattice spin model is given by~\cite{YiSD2009,Mochizuki2012,Buhrandt2013,Tanaka2020},
\begin{eqnarray}
\mathcal {H}=\mathcal{H}_{\rm ex}+\mathcal{H}_{\rm DM}+\mathcal{H}_{\rm Zeeman},
\label{eq:Hamiltonian}
\end{eqnarray}
with
\begin{align}
\label{eq:Hex}
&\mathcal{H}_{\rm ex}=-J \sum_i  \left[
\bm m_i \cdot \bm m_{i+\hat{\bm x}}+\bm m_i \cdot \bm m_{i+\hat{\bm y}} \right],
\\
\label{eq:HDM}
&\mathcal{H}_{\rm DM}=D \sum_i \left[
\hat{\bm y} \cdot (\bm m_i \times \bm m_{i+ \hat{\bm x}})
-\hat{\bm x} \cdot (\bm m_i \times \bm m_{i+ \hat{\bm y}}) \right],
\\
\label{eq:HZeeman}
&\mathcal{H}_{\rm Zeeman}=-\bm B_{\rm ex}\cdot \sum_i \bm m_i.
\end{align}
Here $\bm m_i$ is the normalized classical magnetization vector at site $i$ ($|\bm m_i|$=1). The ferromagnetic exchange interactions and the DM interactions work between the adjacent site pairs $\langle i,j\rangle$ connected by the nearest-neighbor bonds. The parameters of  this lattice spin model is related with those in the above continuum model as $J=2a\mathcal{A}$, $D=\mathcal{D}a^2$, and $\bar{\bm B}_{\rm ext}=\bm B_{\rm ext}M_{\rm s}a^3$. We assume $a$=2.7 nm, which gives $J$=505.5 meV and $D$=136.7 meV with the ratio $D/J$=0.27.  For these parameters, the diameter of skyrmion takes typically $\lambda_{\rm m}$$\sim$90 nm. In the present study, we discuss the skyrmion creation in the ferromagnetic phase on the verge of the phase boundary to the skyrmion-crystal phase. Thus, we fix the $z$-component of the external magnetic field at $B_z$=520 mT.

The magnetization configuration of a skyrmion is characterized by three quantities, i.e.,  the vorticity $n_\Omega(=\pm 1)$, the sign of core magnetization $\zeta_m(=\pm 1)$, and the helicity $\eta$. Using these quantities, the normalized magnetization vectors $\bm m(r, \phi)$ constituting a skyrmion is given in the two-dimensional polar representation as,
\begin{align}
\begin{pmatrix}
m_x \\
m_y \\
m_z 
\end{pmatrix}
=
\begin{pmatrix}
\sin{\Theta(r)}\cos{\Phi(\phi)} \\
\sin{\Theta(r)}\sin{\Phi(\phi)} \\
\cos{\Theta(r)}
\end{pmatrix}.
\end{align}
The spatial profiles of $\Theta(r)$ and $\Phi(\phi)$ are approximately given by,
\begin{align}
&\Theta(r)=\pi\zeta_m \left( \frac{r}{r_{\rm sk}} + \dfrac{\zeta_m - 1}{2} \right), \\
&\Phi(\phi)=n_\Omega \phi + \eta,
\end{align}
where $r_{\rm sk}$ is the radius of the skyrmion. When the magnetic field $\bm B_{\rm ex}=(0, 0, B_z)$ with $B_z>0$ ($B_z<0$) is applied, the sign of core magnetization $\zeta_m$ is $-1$ ($+1$). In the following, we always consider the situation that $B_z>0$ and $\zeta_m=+1$ unless otherwise noted. For the N\'{e}el-type skyrmions, the vorticity $n_\Omega$ is $+1$, and the helicity $\eta$ takes 0 or $\pi$. The helicity $\eta$ corresponds to the rotation sense of magnetizations along the diameter, which is determined by the sign of the DM parameter $D$ in Eq.(\ref{eq:HDM}) [Figs.~\ref{Fig02}(e)-(g)].

The magnetization dynamics driven by the spin-orbit torque due to a vertical spin current are simulated using the Landau-Lifshitz-Gilbert-Slonczewskii (LLGS) equation,
\begin{align}
\frac{d\bm m_i}{dt}
&=\displaystyle -\gamma \bm m_i \times \bm B_i^{\rm eff} 
+\alpha_{\rm G} \bm m_i \times \frac{d\bm m_i}{dt} 
\nonumber \\ 
&+\frac{\gamma\hbar |\theta_{\rm SH}|j_{\rm e}(\bm r_i)}{2eM_{\rm s}d}
\left[ \bm m_i \times (\bm \sigma_i \times \bm m_i) \right].
\label{eq:LLGS}
\end{align}
where $\hbar$ and $\gamma$ are the Planck constant and the gyromagnetic ratio, respectively. The first term describes the precessional motion of magnetization around the local effective field $\bm B_i^{\rm eff}$ , which is calculated by,,
\begin{eqnarray}
\displaystyle
\bm B_i^{\rm eff}=-\frac{1}{M_{\rm s}a^3}\frac{\partial \mathcal{H}}{\partial \bm m_i}.
\end{eqnarray}
The second term is the phenomenologically introduced Gilbert-damping term. The dimensionless coefficient $\alpha_{\rm G}$ is fixed at $\alpha_{\rm G}$=0.04 throughout this work. The third term represents the spin-orbit torque exerted by the vertical spin current with spin polarization $\bm \sigma_i$. Here $d$(=2.7 nm) is the thickness of the ferromagnetic layer, $j_{\rm e}(\bm r_i)$ is the local electric current density, $\theta_{\rm SH}$ is the spin Hall angle, and $e(>0)$ is the elementary charge. Note that although the spin-transfer torque also influences magnetization dynamics, its effect is considerably weak as compared to that of the spin-orbit torque in the multilayer system, and thus we neglect the spin-transfer torque in this study.

The electric current injected to the heavy-metal layer is converted to the perpendicular pure spin current by the spin Hall effect. The conversion efficiency is represented by the spin Hall angle $\theta_{\rm SH}=j_{\rm e}/j_{\rm s}$ with $j_{\rm e}$ and $j_{\rm s}$ being the electric current density and the spin current density, respectively. We adopt a realistic value of $\theta_{\rm SH}$=0.1 in the following simulations.
The spin polarization $\bm \sigma_i$ of the vertical spin current is governed by the direction of electric-current flow. Specifically, the local orientation of $\bm \sigma_i$ should be perpendicular to the local current-density vector $\bm j_{\rm e}(\bm r_i)$ and parallel to the interface plane. Therefore,  $\bm \sigma_i$ is given by,
\begin{eqnarray}
\bm \sigma_i={\rm sgn}(\theta_{\rm SH})
\frac{\bm j_{\rm e}(\bm r_i) \times \bm e_z}{|\bm j_{\rm e}(\bm r_i) \times \bm e_z|}.
\end{eqnarray}
It is assumed that the current-density vector $\bm j_{\rm e}(\bm r)$ is proportional to the local electric field $\bm E(\bm r)=-\nabla \phi(\bm r)$ with $\phi(\bm r)$ being the electric potential, which leads to $\bm j_{\rm e}(\bm r) \propto \nabla \phi(\bm r)$. For a steady current distribution, the conservation law of current, i.e., $\bm \nabla \cdot \bm j_{\rm e}=0$, holds, which eventually leads to the Poisson equation $\Delta \phi(\bm r)$=0. When the stripline is straight with a uniform width and has no structure, the eletric potential $\phi(\bm r)$ is spatially uniform and constant. On the contrary, when the stripline has a notch structure, $\phi(\bm r)$ is no longer uniform but becomes spatially modulated.

To obtain the spatial profile of $\phi(\bm r)$, we solve the Poisson equation using the finite-element method by imposing the following boundary conditions:
\begin{eqnarray}
& &\phi(\bm r)=
\begin{cases}
0         & (\text{at the left edge of sample})\\
\phi_0 & (\text{at the right edge of sample})
\end{cases}
\label{eq:NBC}
\\
& &\frac{\partial \phi}{\partial \bm n}=0
\quad (\text{at the other edges})
\label{eq:DBC}
\end{eqnarray}
The differentiation $\partial \phi/\partial \bm n$ denotes a spatial derivative of the potential $\phi$ at a system boundary along the direction normal to the boundary. With the first two conditions (the Dirichlet boundary condition), the electric current is flowing into the system at the right end, while flowing out from the system at the left end where the electrons are flowing from left to right. The latter condition (the Neumann boundary condition) corresponds to the charge conservation. We employ a nanometric stripline of 2700 nm $\times$ 135 nm $\times$ 2.7 nm (1000$\times$50$\times$1 cells) for the calculation of electric current distribution using the Poisson equation but focus on an area of 1620 $\times$ 135 nm $\times$ 2.7 nm (600$\times$50$\times$1 cells) around the notch structure for the simulation of magnetic dynamics.

In the simulations, we calculate time profiles of the total topological charge $N_{\rm sk}$ to count the number of created skyrmions. This quantity is give by,
\begin{eqnarray}
N_{\rm sk}=\frac{1}{4\pi}
\displaystyle  \sum_i \bm m_i \cdot
(\bm m_{i+\hat{x}} \times \bm m_{i+\hat{x}+\hat{y}})
\label{TopoC}
\end{eqnarray}
Note that the topological number $N_{\rm sk}$ is not zero even in a system without skyrmions before the current injection. This is because the DM interaction induces a winding alignment of the magnetization along the system edges, resulting in a small but finite topological charge. We regard the value obtained after subtracting this edge contribution as the bulk topological charge $N_{\rm sk}$ of the system.

\section{Results and discussion}
\begin{figure}[tbh]
\centering
\includegraphics[scale=1.0]{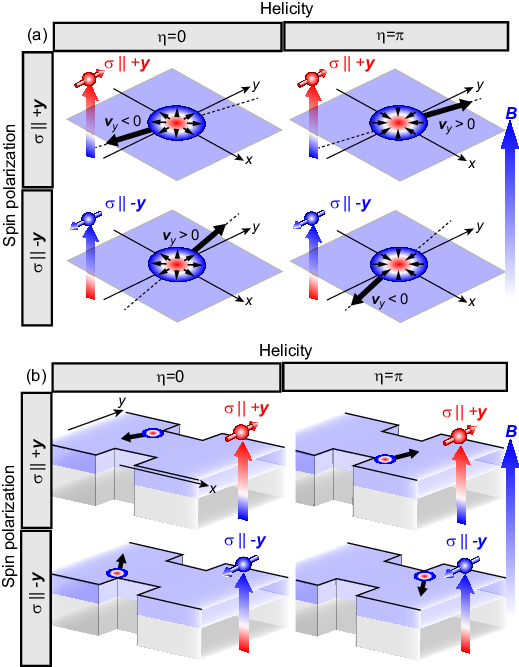}
\caption{(a) Dependence of the moving direction of a N\'eel-type skyrmion driven by spin-orbit torque on the spin polarization $\bm \sigma$($\parallel$$\pm \bm y$) of the vertical spin current and the helicity $\eta$ of the spatial magnetization configuration of skyrmion. The skyrmion moves almost in the $\pm \bm y$ direction (parallel or antiparallel to $\bm \sigma$), while slightly in the $\pm \bm x$ direction (perpendicular to $\bm \sigma$) due to the skyrmion Hall effect. (b) Dependence of the corner of the rectangular notch at which skyrmions are created by spin-orbit torque on $\bm \sigma$ and $\eta$. Because of the dependence in (a), skyrmions can be created only at a specific corner of rectangular notch depending on $\bm \sigma$ and $\eta$.}
\label{Fig03}
\end{figure}
First, the simulations show that the direction of motion of N\'{e}el-type skyrmion driven by spin-orbit torque depends on the helicity $\eta$ ($\eta=0$ or $\eta=\pi$) of the skyrmion magnetization configuration and the spin polarization $\bm \sigma$ ($\bm \sigma$$\parallel$$+\bm y$ or $\bm \sigma$$\parallel$$-\bm y$) of the spin current. Figure~\ref{Fig03}(a) summarizes the direction of motion of the skyrmion for four combinations of helicity $\eta$ and spin polarization $\bm \sigma$. In all cases, the skyrmions move mainly in the direction parallel or antiparallel to the spin polarization $\bm \sigma$, i.e., in the $+y$ or $-y$ direction. In addition, they have a small velocity component in the direction orthogonal to the spin polarization $\bm \sigma$, i.e., in the $+x$ or $-x$ direction, due to the skyrmion Hall effect. This dependence on the direction of motion causes dependence of the creation position, i.e., from which corner of the notch the skyrmion is created when it is created by the spin-orbit torque in a magnetic bilayer heterojunction with a rectangular notch. In other words, for a skyrmion seed created by the local magnetization reversal that occurs at a corner of the notch to grow into a skyrmion, it must successfully leave the corner and move deep inside the ferromagnetic track. Figure~\ref{Fig03}(b) summarizes the dependence of the skyrmion-creation position on the helicity $\eta$ and the spin polarization $\bm \sigma$ clarifed by taking this aspect into account. From this figure, it can be seen that the side (upper side or lower side) of the magnetic nanotrack on which the notch is placed is an important factor to observe the skyrmion creation through the spin-orbit torque in a magnetic bilayer heterojunction.

\begin{figure*}[tbh]
\centering
\includegraphics[scale=1.0]{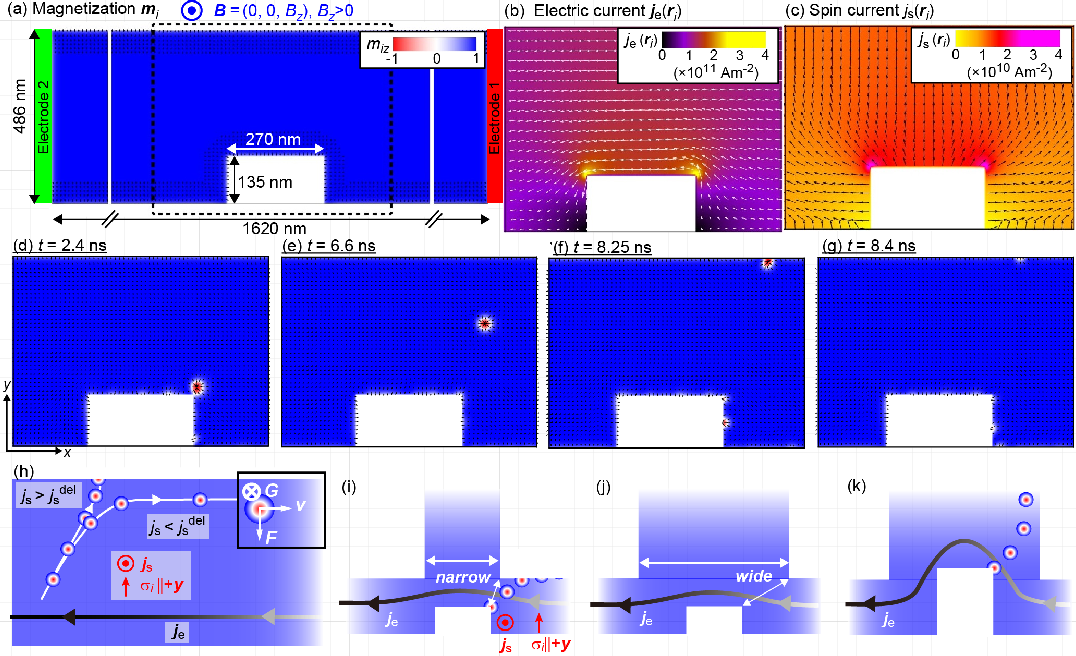}
\caption{(a) Nanotrack system with a rectangular notch used for micromagnetic simulatioins. Two electrodes are set at the left and right ends, and thus the electric current flows in the width ($x$) direction. (b) Spatial profile of the electric current $\bm j_{\rm e}(\rm r_i)$ in the area arround the notch indicated by the dashed rectangle in (a). The arrows and colors represent the local diretion $\bm j_{\rm e}(\rm r_i)/j_{\rm e}(\rm r_i)$ and density $j_{\rm e}(\rm r_i)$. (c) Spatial profile of the vertical spin current $\bm j_{\rm s}(\rm r_i)$ where the arrows and colors represent the local spin polarization $\bm \sigma_i$ and density $j_{\rm s}(\rm r_i)$. (d)-(g) Snapshots of the skyrmion creation and motion driven by the spin-orbit torque at selected moments when $\sigma$$\parallel$$+\bm y$, $\eta=\pi$ and $j_{\rm s}=1.7 \times 10^{11}$ Am$^{-2}$. (h) Distinct behaviors of a skyrmion after creation, which moves towards the upper edge. The skyrmion collides with the upper edge and is absorbed when the current density $j_{\rm e}$ is greater than the threshold value $j_{\rm e}^{\rm del}$ for the skyrmion annihilation. On the contrary, it moves in the $+\bm x$ direction along the edge when $j_{\rm e}$ is smaller than $j_{\rm e}^{\rm del}$. (i)-(k) Three types of (inverted) T-shaped systems with a rectangular notch, i.e., (i) system with a narrower vertical track and a notch shorter in height, (j) system with a wider vertical track and a notch shorter in height, and (k) system with a wider vertical track and a taller notch.}
\label{Fig04}
\end{figure*}
To discuss the difficulty of skyrmion creation by spin-orbit torque, we argue the results of numerical simulations for a certain situation. We consider a nanotrack of 1620 nm $\times$ 486 nm $\times$ 2.7 nm (600$\times$180$\times$1 cells) with a rectangular notch of 270 nm $\times$ 135 nm $\times$ 2.7 nm (100$\times$50$\times$1 cells) [Fig.~\ref{Fig04}(a)]. A perpendicular magnetic field $\bm B=(0,0,B_z)$ with $B_z$=520 mT is applied to the sample plane, and thus the magnetizations point upward in the initial state. An electric current with $j_{\rm e}=1.7 \times 10^{12}$ Am$^{-2}$ is injected into the heavy-metal layer of the heterojunction from Electrode 1 and is flow out from Electrode 2, which are attached, respectively, at right and left ends of the nanotrack. We simulate the skyrmion creation with spin-orbit torque in this situation. Figure~\ref{Fig04}(b) shows the spatial distribution of the current density $j_{\rm e}(\bm r_i)$ (colors) and the direction $\bm j_{\rm e}(\bm r_i)/j_{\rm e}(\bm r_i)$ (arrows) of electric current in the region enclosed by the dashed rectangle in Fig.~\ref{Fig04}(a). This figure shows that the current density is locally enhanced at the two corners of the rectangular notch. On the other hand, Fig.~\ref{Fig04}(c) shows the spatial distribution of the intensity $j_{\rm s}(\bm r_i)$ (color) and the spin polarization $\bm \sigma_i$ (arrows) of the vertical spin current generated from this electric current by the spin Hall effect from this electric current. 

Figures~\ref{Fig04}(d)-(g) are snapshots of the spatiotemporal dynamics of magnetization obtained in the simulation in this situation. First, it can be seen that a local magnetization reversal occurs at the upper right corner of the rectangular notch, leading to a skyrmion creation. This is induced by the locally enhanced spin current at the corner. Furthermore, the created skyrmion leaves the corner and move toward the upper edge of the nanotrack. The skyrmion moves mainly in the longitudinal direction (the width direction) or the $+y$ direction, although it also has a small velocity component in the transverse direction (the length direction) or $+x$ direction due to the skyrmion Hall effect. The skyrmion reaches and collides with the upper edge, and, eventually, it is absorbed by the edge and disappear. Figure~\ref{Fig04}(h) shows different behaviors of the skyrmion after it collides with the upper edge depending on the spin current density $j_{\rm s}$. If $j_{\rm s}$ is smaller than a certain threshold $j_{\rm s}^{\rm del}$, i.e., $j_{\rm s}<j_{\rm s}^{\rm del}$, the skyrmion that reaches the upper edge continues to move in the length direction of the nanotrack, being suffered from a repulsive potential force from the edge. On the contrary, if $j_{\rm s}$ is larger than the threshold, i.e., $j_{\rm s}>j_{\rm s}^{\rm del}$, the skyrmion that collides with the upper edge is absorbed by the edge and disappears. Since the threshold spin current density $j_{\rm s}^{\rm cre}$ required to create a skyrmion is one or two orders of magnitude larger than $j_{\rm s}^{\rm del}$, a skyrmion created in this way will necessarily be absorbed and disappear unless the application of electric current is stopped immediately after its creation.

The problem lying in the skyrmion creation with spin-orbit torque in a notched nanotrack system is that the N\'{e}el-type skyrmion created at the notch moves in a direction perpendicular to the electric current flowing in the nanotrack. A simple solution to this problem is to introduce an additional track perpendicular to the notched nanotrack or the flowing electric current which works as a channel for the moving skyrmions after created. This idea actually works as will be demonstrated in the following. However, the width of the attached vertical track and the height of the notch must be carefully designed. For example, if the width of the vertical track is narrow as shown in Fig.~\ref{Fig04}(i), the skyrmion created at the notch corner will not enter the vertical track properly. This is because the skyrmions do not move straight in the width direction due to the skyrmion Hall effect. However, this does not mean that the width of the vertical track should simply be widened. For example, if a wider vertical track is introduced as shown in Fig.~\ref{Fig04}(j), the electric current density and the resulting spin current density at the corner of the notch cannot be large enough to induce the local magnetization reversal required for the skyrmion creation. The ideal sample geometry to realize the above idea is to introduce a wide vertical track and, at the same time, increase the notch height as shown in Fig.~\ref{Fig04}(k). This will ensure that the sufficiently increased current density at the strongly constricted area will create skyrmions via the magnetization reversal and that the created skyrmions will be captured by the wide vertical track.

\begin{figure*}
\centering
\includegraphics[scale=1.0]{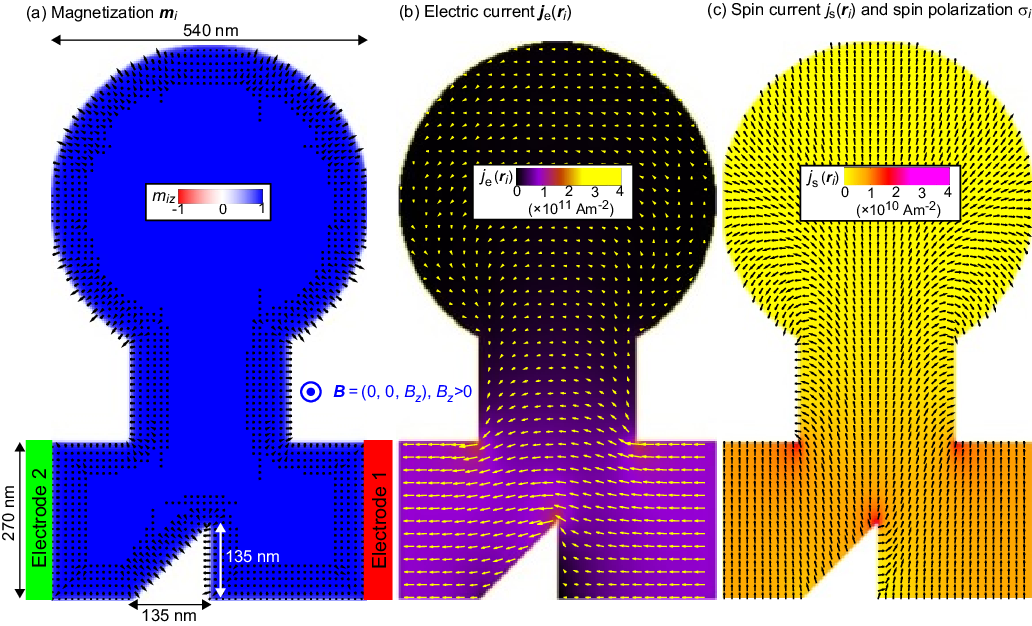}
\caption{Spatial profiles of (a) magnetizations $\bm m_i$, (b) electric current $\bm j_{\rm e}(\bm r_i)$, and (c) vertical spin current $\bm j_{\rm s}(\bm r_i)$ in the inverted T-shaped system with a triangular notch and a circular chamber. The colors indicate the (a) out-of-plane components $m_{zi}$, (b) electric current density $j_{\rm e}(\bm r_i)$, and (c) spin current density $j_{\rm s}(\bm r_i)$. On the other hand, the arrows indicate the (a) in-plane components $(m_{xi}, m_{yi})$, (b) electric current direction $\bm j_{\rm e}(\bm r_i)/j_{\rm e}(\bm r_i)$, and (c) spin polarization $\bm \sigma_i$ of the vertical spin current.}
\label{Fig05}
\end{figure*}
We refer to the system with vertical nanotracks attached to the horizontal nanotrack introduced above as a T-shaped system. Here we discuss the results of simulations for the skyrmion creation with spin-orbit torque when a triangular notch is introduced in the T-shaped system. We performed simulations in both the rectangular notch case and the triangular notch case and found that skyrmions can be created and injected to the vertical track in both cases. However, in the case of the rectangular notch, we found that the sample-shape parameters need to be tuned somewhat precisely to create skyrmions stably from the notch corner. On the other hand, in the case of the triangular notch, it was found that skyrmions can be stably created from the apex of the triangular notch without adjusting the sample-shape parameters so precisely. We will now present the results for the triangular notch case. We consider a T-shaped system compsed a horizontal nanotrack with a width of 270 nm (100 cells) and a vertical nanotrack with a width of 270 nm (100 cells). A right-angled triangular notch with a base of 135 nm (50 cells) and a height of 135 nm (50 cells) is introduced on the horizontal nanotrack as shown in Fig.~\ref{Fig05}(a). At the end of the vertical nanotrack, a circular chamber region was placed to accumulate the created skyrmions. An electric current is injected from Electrode 1 and is flow out from Electrode 2. Figure~\ref{Fig05}(b) shows the spatial distribution of the current density $j_{\rm e}(\bm r_i)$ (colors) and direction $\bm j_{\rm e}(\bm r_i)/j_{\rm e}(\bm r_i)$ (arrows) of the electric current flowing in the heavy-metal layer in the magnetic bilayer heterojunction. On the other hand, Fig.~\ref{Fig05}(c) shows the spatial distribution of the intensity $j_{\rm s}(\bm r_i)$ (color) and the spin polarization $\bm \sigma_i$ (arrows) of the vertical spin current injected into the ferromagnetic layer. 

\begin{figure*}
\centering
\includegraphics[scale=1.0]{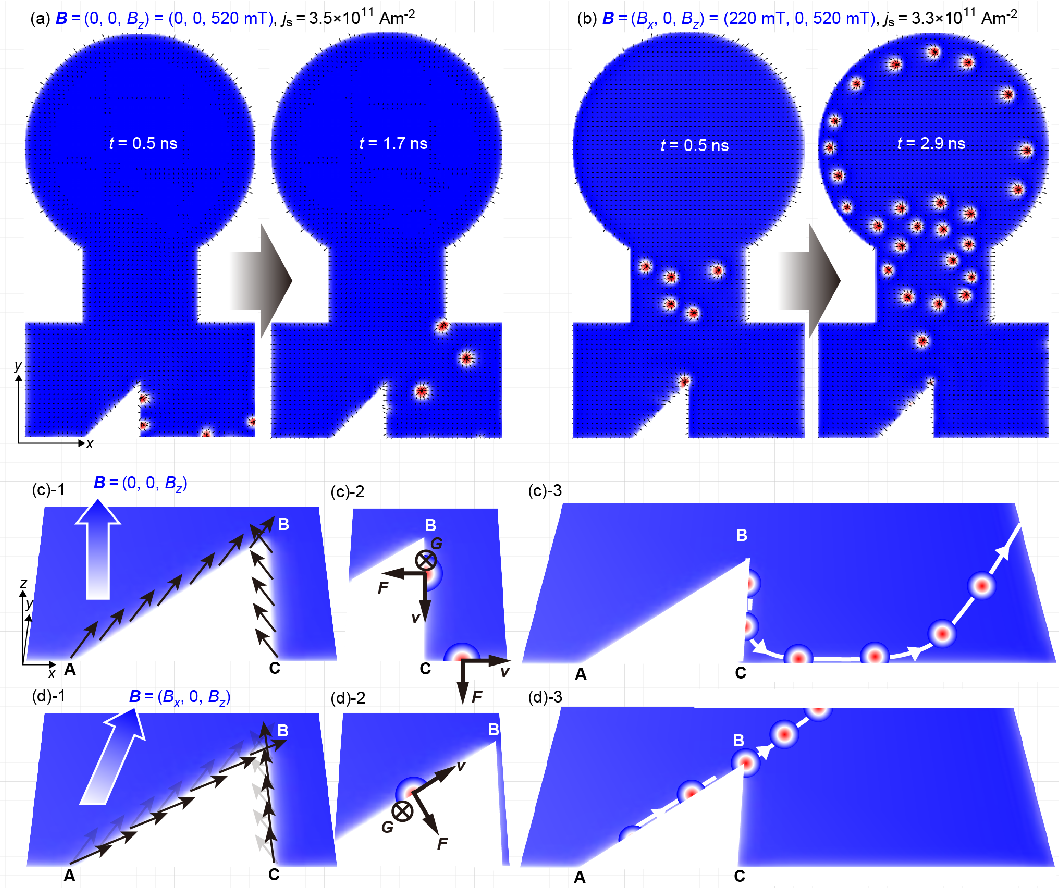}
\caption{Skyrmion creations with spin-orbit torque in the (inverted) T-shaped system with a right-triangle notch. (a) When the external magnetic field is perpendicular with $B_x=0$, $B_y=0$ and $B_z$=520 mT, skyrmion seeds are created at the vertical side of the right-triangle notch. They eventually disappear being absorbed by sample edges and cannot survive. (b) When the external magnetic field is tilted in the $x$ direction with $B_x$=220 mT, $B_y=0$ and $B_z$=520 mT, skyrmion seeds are created at the oblique side of the right-triangle notch. They survive without being absorbed by sample edges and are accumulated in the circular chamber located above the notch. (c) Magnetization configuration on edges of the triangular notch (left panel), relationship among the gyromagnetic vector $\bm G$, the attractive force from the edge $\bm F$ and the skyrmion velocity $\bm v$ (middle panel), and trajectory of a created skyrmion seed (right panel) when the external magnetic field is perpendicular. (d) Those when the external magnetic field is tilted in the $+x$ direction.}
\label{Fig06}
\end{figure*}
Figures~\ref{Fig06}(a) and (b) show the results obtained from the simulations on the skyrmion creation. First, Fig.~\ref{Fig06}(a) shows the results for the case where the external magnetic field $\bm B$ is perpendicular to the sample plane, i.e., $\bm B=(B_x, 0, B_z)$ with $B_x=0$. Here we take $B_z$=520 mT to set the system inside the ferromagnetic phase close to the phase boundary to the skyrmion crystal phase. The spin current density at the electrode is fixed at $j_{\rm s}=3.5 \times 10^{11}$ Am$^{-2}$. The left panel of Fig.~\ref{Fig06}(a) shows that local magnetization reversal due to the spin-orbit torque occurs on the perpendicular side of the right-triangle notch. We also find that the created skyrmion seeds or half-skyrmions move to the right along the sample edge. The skyrmion seeds continue to move along the sample edge and leave the edge to be real skyrmions when they are far away from the notch. At this time, because they are far away from the area below the vertical nanotrack, most of the created skyrmions do not enter the vertical track but instead collide with the upper edge or corner and disappear. One may think that the skyrmions can enter the vertical notch if we put the vertical track away from the notch. However, in this way, we cannot realize the sufficient enhancement of the current density required for the local magnetization reversal.

We find that this problem can be solved by applying a tilted magnetic field. Figure~\ref{Fig06}(b) shows the result when the external magnetic field is slightly tilted in the x direction from the perpendicular direction, i.e., $\bm B=(B_x, 0, B_z)$ with $B_x \ne 0$. In this case, the local magnetization reversal due to spin-orbit torque occurs at the hypotenuse or the oblique side of the right-triangle notch. Furthermore, skyrmion seeds or half-skyrmions with reversed magnetization run up the oblique side. They leave the notch and go inside the nanotrack. Above the apex of the triangular notch, there is the vertical track. The created skyrmions are successfully injected into the vertical track.

This difference arises from the different arrangement of magnetizations along the triangle notch depending on whether the magnetic field is perpendicular or tilted. The magnetizations inside the nanotrack are oriented mostly parallel to the applied magnetic field $\bm B$, while the magnetizations at sample edges have large in-plane components due to the imbalance of the Dzyaloshinskii-Moriya interaction. Specifically, there are neighboring magnetizations coupled via the exchange and Dzyaloshinskii-Moriya interactions on one side, but no magnetization on the opposite side for the magnetizations at the sample edges. The spatial arrangement of the in-plane magnetization components along the notch is an important factor that determines possibility and location of the local magnetization reversal by the spin-orbit torque. When the magnetic field is perpendicular, the in-plane magnetization components along the notch are arranged as in Fig.~\ref{Fig06}(c)-1. In this case, the magnetization reversal tends to occur on the perpendicular side of the notch. On the other hand, when the magnetic field is tilted in the $+x$ direction, the in-plane magnetization components along the notch are arranged as in Fig.~\ref{Fig06}(d)-1. In this case, the magnetization reversal tends to occur on the oblique side of the notch. The half-skyrmions as seeds of skyrmions which appear as a consequence of the magnetization reversal are suffered from an attractive potential force $\bm F$ from the edge [Figs.~\ref{Fig06}(c)-2 and (d)-2]. According to the Thiele equation~\cite{Thiele1973}, the skyrmion seeds acquire a rightward velocity along the edge. Specifically, the velocity $\bm v$ of the topological magnetic texture with a finite topological number $\mathcal{G}$ under the influence of potential force $\bm F=-\nabla V$ is given by the equation,
\begin{align}
\bm G \times \bm v=\bm F
\end{align}
where $\bm G \equiv (0,0,\mathcal{G})$ is the gyromagnetic vector. In the continuum limit, the topological invariant $\mathcal{G}$ is given by,
\begin{align}
\mathcal{G} \equiv \int_{\rm UC} \bm m \cdot \left(
\frac{\partial \bm m}{\partial x}\times \frac{\partial \bm m}{\partial y} \right) \;dxdy.
\end{align}
Because $\mathcal{G}=4\pi \zeta_{\rm m}$ for a skyrmion and $\mathcal{G}=2\pi \zeta_{\rm m}$ for a half-skyrmion with $\zeta_{\rm m}=\pm 1$ being the sign of the core magnetization,  $\mathcal{G}=-4\pi$ and $\mathcal{G}=-2\pi$ for the skyrmion and the half-skyrmion with $\zeta_{\rm m}=-1$, respectively, in the present case. Consequently, the created skyrmion seeds and skyrmions move as shown in Fig.~\ref{Fig06}(c)-3 when the magnetic field is perpendicular, while as shown in Fig.~\ref{Fig06}(d)-3 when the magnetic field is tilted, as observed in the simulations.

\begin{figure}[tbh]
\centering
\includegraphics[scale=1.0]{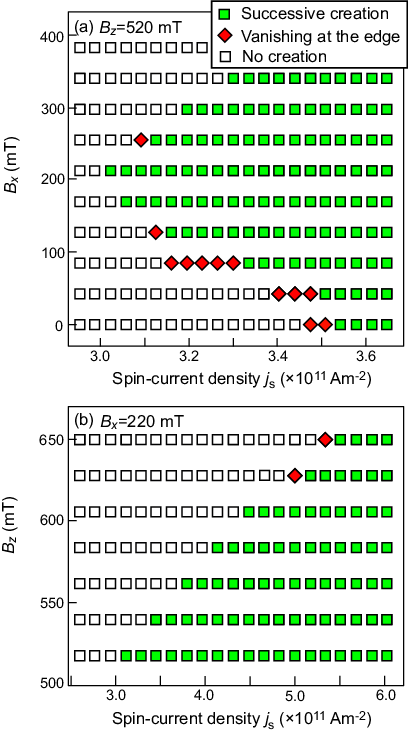}
\caption{Diagrams of skyrmion creation with spin-orbit torque in the system shown in Fig.\ref{Fig06} under application of the external magnetic field $\bm B \equiv (B_x, 0, B_z)$. The horizonal axis represents the spin current density $j_{\rm s}$. (a) Diagram where the vertical axis represents $B_x$ when $B_z$ is fixed at 520 mT. (b) Diagram where the vertical axis represents $B_z$ when $B_x$ is fixed at 220 mT.}
\label{Fig07}
\end{figure}
The diagrams in Fig.~\ref{Fig07} summarize the dependence of skyrmion creation on the spin current density $j_{\rm s}$ and the external magnetic field $\bm B=(B_x,0,B_z)$ in the T-shaped magnetic bilayer heterojunction with a right triangle notch in Fig.~\ref{Fig05}. Note that when the magnetic field has only a perpendicular component $B_z(\ne 0)$ with zero in-plane components $B_x=0$ and $B_y=0$, the critical field strength for the phase boundary bwteen the skyrmion crystal and ferromagnetic phases is $B_z$=490 mT for the material parameters used in this study. First, Fig.~\ref{Fig07}(a) is a diagram in the plane of $j_{\rm s}$ and $B_x$ when $B_z$=520 mT. It is important to note that when the tilting of the magnetic field is moderate with $B_x$$\approx$220 mT, the spin current density necessary for skyrmion creation takes a minimum value. On the other hand, Fig.~\ref{Fig07}(b) is a diagram in the plane of $j_{\rm s}$ and $B_z$ when $B_x$=220 mT. This diagram shows that as the system goes deeper into the ferromagnetic phase leaving from the phase boundary to the skyrmion crystal phase with increasing $B_z$, the spin current density necessary for skyrmion creation increases. These two diagrams indicate that in order to efficiently create skyrmions with spin-orbit torque in the magnetic bilayer heterojunction system, a perpendicular magnetic field $B_z$ should be applied so that the system is located in the ferromagnetic phase near the skyrmion crystal phase, while an in-plane magnetic field $B_x$ should be approximately of one-half to one-third of $B_z$.

\begin{figure*}[tbh]
\centering
\includegraphics[scale=1.0]{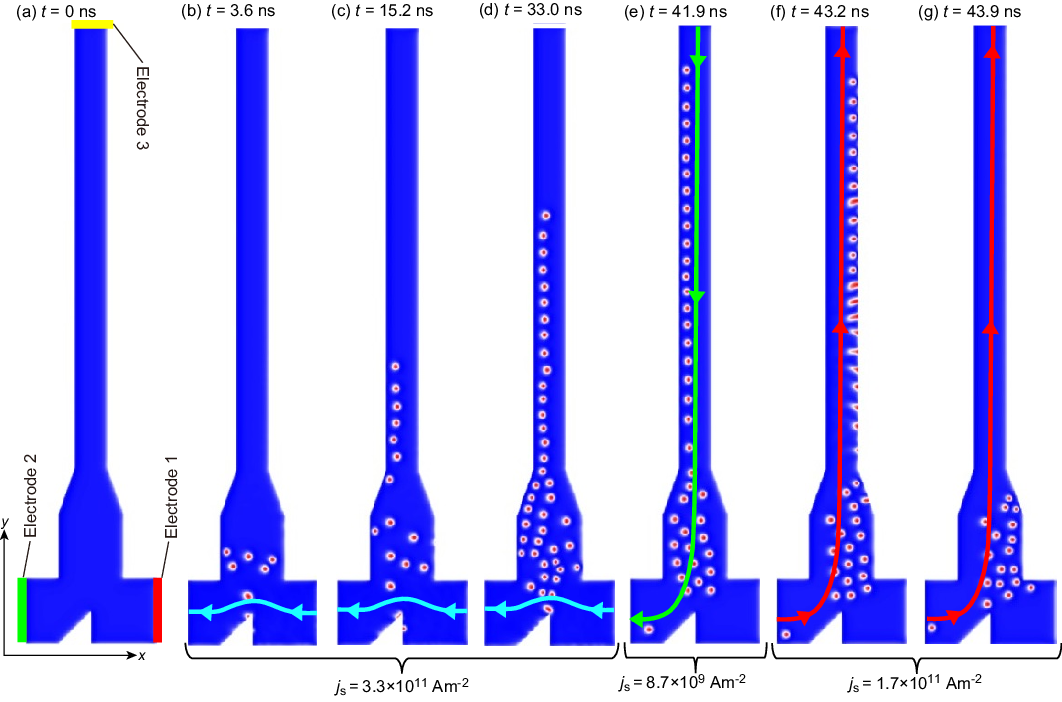}
\caption{Snapshots of the skyrmion operation (creation, driving, and deletion) with spin-orbit torque in the T-shaped magnetic bilayer heterojunction system with a right-triangle notch at selected moments. (a) Three electrodes 1, 2, and 3 are attached at respective terminals. A tilted magnetic field $\bm B \equiv (B_x, 0, B_z)$ is applied where $B_x$=220 mT and $B_z$=520 mT. (b)-(d) Creation operation by injecting a large electric current from electrode 1 which flows to electrode 2 in the horizontal track. Skyrmions are created successively at the triangular notch and move to the vertical track. (e) Driving operation by injecting a small electric current from electrode 3 which flows to electrode 2 mainly in the vertical track. Skyrmions move upwards in the vertical track, while no skyrmions are created at the notch because the current density is much smaller than the threshold value for the skyrmion creation. (f),~(g) Deletion operation by injecting a large electric current from electrode 2 which flows to electrode 3 mainly in the vertical track. All the skyrmions in the vertical track move to the right wall of the track and disappear being absorbed by the wall. The skyrmion creation does not occur at the notch because the current direction is opposite to that for the skyrmion creation. The spin current density $j_{\rm s}$ measured at electrode 2  at each moment is (b)-(d) $3.3 \times 10^{11}$ Am$^{-2}$, (e) $8.7 \times 10^{9}$ Am$^{-2}$, and (f),~(g) $1.7 \times 10^{11}$ Am$^{-2}$.} 
\label{Fig08}
\end{figure*}
Finally, we demonstrate the operation of skyrmions with spin-orbit torque in a T-shaped system with a right-triangle notch [Figs.~\ref{Fig08}(a)-(g)]. As shown in Fig.~\ref{Fig08}(a), three electrodes 1, 2, and 3 are attached at each end. In addition, a tilted external magnetic field $\bm B \equiv (B_x, 0, B_z)$ with $B_x$=220 m\t and $B_z$=520 mT is applied. First, an electric current is applied in the horizontal direction from electrode 1 to electrode 2 as shown in Fig.~\ref{Fig08}(b). At this time, the spin current density $j_{\rm s}(=\theta_{\rm SH}j_{\rm e})$ at electrode 2 is set at $3.3 \times 10^{11}$ Am$^{-2}$. In this situation, skyrmions are created from the triangular notch. As the electric current continues to be applied, skyrmions are successively created at equivalent time interval. The created skyrmions move upward (in the $+y$ direction) and enter the vertical track [Figs.~\ref{Fig08}(b)-(d)]. Then, we apply a voltage between electrodes 2 and 3 to inject an electric current flowing from the vertical track to the left side of the horizontal track as shown in Fig.~\ref{Fig08}(e). In this situation, the skyrmions in the vertical track form a line with equally spaced intervals and move in the $+y$ direction at a steady speed of 70 ms$^{-1}$. Here the spin current density $j_{\rm s}$ at electrode 2 is set at $8.7 \times 10^{9}$ Am$^{-2}$. Because this spin current density is sufficiently small compared to the threshold density $j_{\rm s}^{\rm cre}$ for the skyrmion creation, no skirmions are created at the triangular notch. Thereby, a pure driving operation of skyrmions is realized. Subsequently, we apply an electric current in the opposite direction to that in Fig.~\ref{Fig08}(e). Specifically, as shown in Fig.~\ref{Fig08}(f), a voltage is applied between electrodes 2 and 3 to inject an electric current flowing from the left side of the horizontal track to the vertical track. Here the spin current density $j_{\rm s}$ at electrode 2 is set at $1.7 \times 10^{11}$ Am$^{-2}$. In this situation, all the skyrmions in the vertical track move toward the right wall and disappear through being absorbed by the wall [Figs.~\ref{Fig08}(f) and (g)]. It should be noted that despite the application of a very large spin current, no skyrmions are created at the notch because the current direction is opposite to that for which the skyrmion creation occurs. Therefore, a pure deleting operation of skyrmions is achieved.

\begin{figure}
\centering
\includegraphics[scale=1.0]{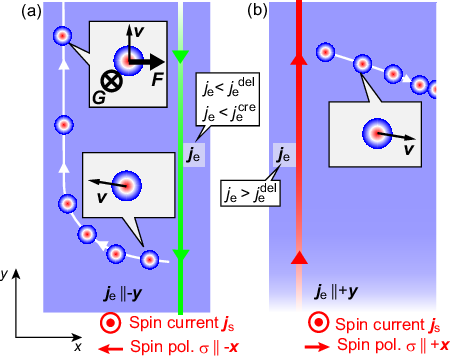}
\caption{(a) Schematics of the relations among the electric current $\bm j_{\rm e}$, the spin current $\bm j_{\rm s}$ and the skyrmion motion driven by the spin-orbit torque in Fig.~\ref{Fig08}(e). (b) Schematics of those in Figs.~\ref{Fig08}(f) and (g).}
\label{Fig09}
\end{figure}
Figure~\ref{Fig09}(a) shows the electric current $\bm j_{\rm e}$ and the skyrmion motion in the vertical track in the case of Fig.~\ref{Fig08}(e). On the other hand, Fig.~\ref{Fig09}(b) shows those in the case of Figs.~\ref{Fig08}(f) and (g). The skyrmion considered here is a N\'{e}el type with a helicity $\eta=\pi$ and a negative core magnetization $\zeta_{\rm m}=-1$. In Fig.~\ref{Fig09}(a), the current $\bm j_{\rm e}$ flows in the $-y$ direction. In this case, the skyrmion initially moves to the left and collides with the left wall of the track. Upon the collision, the skyrmion is suffered from a rightward force $\bm F$ due to the repulsive potential from the wall. Consequently, it moves forward in the $+y$ direction in accord with the Thiele equation. At this time, if the current density $j_{\rm e}$ is large, the skyrmion might be absorbed by the wall and disappear. However, because $j_{\rm e}$ is smaller than a threshold for the skyrmion annihilation by the wall in the present case, the skyrmion does not disappear and continues to move along the track. In addition, because $j_{\rm e}$ is smaller than a threshold for the skyrmion creation as well, the skyrmion creation dones not occur at the notch. On the other hand, in Fig.~\ref{Fig09}(b), the current $\bm j_{\rm e}$ flows in the $+y$ direction, and the skyrmion moves to the right and collide with the right wall. In this case, because $j_{\rm e}$ is greater than the threshold for the skyrmion annihilation at the wall, the skyrmion that collides with the wall is absorbed and disappear. In addition, although $j_{\rm e}$ is sufficiently large, the skyrmion creation does not occur at the notch because the current $\bm j_{\rm e}$ flows in the opposite direction to that for which skyrmions are created at the notch. Thus, by using the T-shaped device with a right-triangle notch, the operations of creating, driving, and deleting skyrmions by electric current through spin-orbit torque can be achieved.

\section{Summary}
In summary, we have theoretically demonstrated that N\'{e}el-type skyrmions can be successfully created by the electric current injection to the magnetic bilayer heterojunction system via the spin-orbit torque mechanism in a controlled manner. The electric current injected to the heavy-metal layer is converted into a perpendicular pure spin current via the spin Hall effect caused by the interfacial spin-orbit interaction, which exerts magnetic toque on the magnetization in the ferromagnetic layer. To take advantage of this mechanism for the skyrmion creation, we have proposed two ingenuities. One is the notch structure, and the other is the $T$-shaped device geometry. The notch structure works to locally enhance the spin current density and to relax the topological constraints, enabling a gradual rotation and resulting local reversal of magnetization to form a skyrmion core. The $T$-shaped nanotrack system is suitable for realizing a flow of the created skyrmions to let them survive. The N\'{e}el-type skyrmions driven by the spin-orbit torque move in a direction perpendicular to the injected current, and, therefore, they should collide with an edge of the track and disappear in the case of simple straight nanotracks. In the $T$-shaped system, the electric current for the skyrmion creation can be applied perpendicular to the track in which the created skyrmions flow. We have further demonstrated that skyrmions can be created, driven, and deleted by electric currents through spin-orbit torque in the $T$-shaped device with three electrodes at each terminal.Our proposals will contribute significantly to fundamental technologies required for memory applicaions of magnetic skyrmions.

In summary, we have theoretically demonstrated that N'{e}el-type skyrmions can be successfully created in a magnetic bilayer heterojunction system through electric current injection, utilizing the spin-orbit torque mechanism in a controlled manner. The electric current injected into the heavy-metal layer is converted into a pure spin current perpendicular to the layer via the spin Hall effect caused by interfacial spin-orbit interaction. This spin current exerts a magnetic torque on the magnetization in the ferromagnetic layer on the heavy-metal layer. To realize skyrmion creation through this mechanism, we have proposed two structural innovations, i.e., a notch structure and a $T$-shaped device geometry. The notch structure locally enhances the spin current density and relaxes the topological constraints, enabling gradual magnetization rotation and localized reversal that leads to skyrmion core formation. The $T$-shaped nanotrack system is designed to facilitate the motion and survival of the generated skyrmions. Since N'{e}el-type skyrmions driven by spin-orbit torque move perpendicular to the direction of the injected current, they tend to collide with a longitudinal edge of the nanotrack and vanish. However, in the $T$-shaped device, the current for skyrmion creation is applied perpendicular to the track along which the skyrmions flow, allowing for their effective transport. Furthermore, we have demonstrated that skyrmions can be created, driven, and deleted by electric currents via spin-orbit torque in the $T$-shaped device equipped with an electrode at each terminal. Our proposed method offers valuable contributions to the development of fundamental technologies necessary for memory applications of magnetic skyrmions.

\section{Acknowledgment}
MM thanks Shiho Nakamura and Xiuzhen Yu for fruitful discussions. 
This work was supported by 
Japan Society for the Promotion of Science KAKENHI (Grant No.~20H00337, No.~24H02231 and No.~25H00611), 
CREST, the Japan Science and Technology Agency (Grant No.~JPMJCR20T1), 
Waseda University Grant for Special Research Projects (Project No.~2022C-139 and No.~2025C-133),
and
the cooperation of organization between Kioxia Corporation and Waseda University.

\section{Data Availability Statement}
The data that support the findings of this study are available from the corresponding author upon reasonable request.

\end{document}